\documentclass[12pt,preprint]{aastex}
\begin{document}

\title{Gravitational Lensing by a Compound Population of Halos:\\
Standard Models}

\author{Li-Xin Li\altaffilmark{a,1} and Jeremiah P. Ostriker\altaffilmark{b}}
\affil{$^{a}$Harvard-Smithsonian Center for Astrophysics, Cambridge,
MA 02138, USA} \email{lli@cfa.harvard.edu}
\affil{$^{b}$Institute of Astronomy, University of Cambridge, Madingley Road,\\
Cambridge CB3 0HA,
UK} \email{jpo@ast.cam.ac.uk}

\altaffiltext{1}{Chandra Fellow}

\begin{abstract}
Based on observed rotation curves of galaxies and theoretical simulations 
of dark matter halos, there are reasons for believing that at least three 
different types of dark matter halos exist in the Universe classified by 
their masses $M$
and the inner slope of mass density $-\alpha\,$: Population A (galaxies):
$10^{10} h^{-1} M_\odot \la M \la 2\times 10^{13} h^{-1} M_\odot$, $\alpha 
\approx 2$; Population B (cluster halos): $M \ga 2\times 10^{13} h^{-1} 
M_\odot$, $\alpha \approx 1.3$; and Population C (dwarf halos): $M \la 10^{10} 
h^{-1} M_\odot$, $\alpha \approx 1.3$. In this paper we calculate the 
lensing probability produced by such a compound population of dark halos, 
for both image separation and time delay, assuming that the mass function
of halos is given by the Press-Schechter function and the Universe is
described by an LCDM, OCDM, or SCDM model. The LCDM model is normalized to the
{\em WMAP} observations, OCDM and SCDM models are normalized to the 
abundance of rich clusters. We compare the predictions of the
different cosmological models with observational data and show that,
both LCDM and OCDM models are marginally consistent with the current
available data, but the SCDM model is ruled out. The fit 
of the compound model to the observed correlation between splitting angle
and time delay is excellent but the fit to the number vs splitting
angle relation is only adequate using the small number of sources in
the objective JVAS/CLASS survey. A larger survey of the same type would 
have great power in discriminating among cosmological models. Furthermore,
population C in an LCDM model has a unique signature in the time domain,
an additional peak at $\sim 3$ seconds potentially observable in GRBs,
which makes it distinguishable from variants of CDM scenarios, such as 
warm dark matter, repulsive dark matter, or collisional dark matter. 
For image separations greater than 10 arcseconds the differently
normalized LCDM models predict significantly different lensing probabilities
affording an additional lever to break the degeneracies in the CMB 
determination of cosmological parameters.
\end{abstract}


\section{Introduction}

It is well known that gravitational lensing is a powerful tool for 
directly probing the structure and distribution of dark matter in the 
Universe \citep[and references therein]{tur84,sch92,bar01,cou02}. By 
comparing the number of lenses found in a survey of remote sources 
(e.g., quasars, radio galaxies, or high redshift type Ia supernova) to 
theoretical predictions, we should be able to deduce the quantity of
dark matter in the Universe and how it is distributed 
\citep[henceforth LO02; Gladders et al. 2003]{tur90,wam95,por00,kee01a,kee01,li02}. 
The joint observations of gravitational lensing, high redshift type Ia 
supernova, cosmic microwave background (CMB), and cluster abundances 
constrain the Universe to be in all likelihood
flat and accelerating, with the present mass density being composed of
about 70\% cosmological constant (or dark energy), 26\% dark matter, and 
4\% ordinary matter \citep{ost95,bah99,wan00,pea01,efs02,yam02,spe03}.

However, the lensing cross-section (and thus the lensing probability)
is found to be extremely sensitive to the inner density profile of lenses
(Keeton \& Madau 2001; Wyithe, Turner, \& Spergel 2001; LO02). For 
example, with fixed total mass, when the 
inner slope of the density profile, $- \alpha$, changes from $-1$ [the 
NFW case \citep{nav96,nav97}] to $-2$ [the singular isothermal sphere 
(SIS) case \citep{got74,tur84}] while maintaining the same mass density
in lenses, the integral lensing probability 
increases by more than two orders of magnitudes for the flat model of 
the Universe (LO02). Therefore, lensing also sensitively probes small scale 
structure. This complicates matters and renders it is hazardous to use 
observed lensing statistics to draw inferences with regard to cosmology 
before determining the sensitivity to other factors.

In LO02, we have shown that in order to explain the observed
numbers of lenses found in the JVAS/CLASS survey, at least two populations 
of dark halos must exist in nature. One population, which corresponds to
normal galaxies, has masses $\la 10^{13} M_\odot$ and a steep inner density 
profile ($\alpha \approx 2$, i.e. SIS) presumably determined by the 
distribution of baryonic material in the inner parts of 
galaxies;\footnote{Strictly speaking, the mass density profile of
galaxies is the combination of the density profile of hark halos and that
of baryonic material at the centers. For brevity, in the paper we still
call it the mass density of halos, though we mean the sum of the mass
density of dark halos and that of baryonic material whenever we refer to
galaxies.}  the 
other one, which corresponds to groups or clusters of galaxies, has 
masses $\ga 10^{13} M_\odot$ and a shallow inner density profile 
($\alpha \la 1.4$, i.e. similar to NFW). A similar conclusion has been 
obtained by \citet{por00} for explaining the number of lenses found in the 
CASTLES survey. These results are consistent with the theoretical studies 
on the cooling of massive gas clouds: there is a 
critical mass of halos $\sim 10^{13} M_\odot$ below which cooling of
the corresponding baryonic component will lead to concentration of 
the baryons to the inner parts of the mass profile \citep{ree77,blu86}.  

In this paper we investigate the lensing statistics produced by a 
compound population of halos. We assume that there are three populations of 
halos in the Universe: 

\noindent
--- Population A: $10^{10} h^{-1} M_\odot < M < 2\times 10^{13} h^{-1} 
M_\odot$, $\alpha = 2$ (SIS); \\
--- Population B: $M > 2\times 10^{13} h^{-1} M_\odot$, $\alpha = 1.3$ [GNFW 
\citep[generalized NFW,][]{zha96}]; \\
--- Population C: $M < 10^{10} h^{-1} M_\odot$, $\alpha = 1.3$ (GNFW), 

\noindent
where $h$ is the Hubble constant in units of 100 km s$^{-1}$ Mpc$^{-1}$. 
Population A corresponds to spiral and elliptical galaxies, whose centers
are dominated by baryonic matter. Population B corresponds to groups or
clusters of galaxies, whose centers are dominated by dark matter. 
Population C corresponds to dwarf galaxies or subgalactic objects, 
whose centers lack baryons due to feedback processes such as
supernova explosions, stellar winds, and photoionizations 
\citep{kau93,som99,ben02,ma03}, and so are also dominated by dark matter. 
We adopt an inner slope
for the dark matter halos of $\alpha = 1.3$ consistent with the value
$1.3\pm 0.2$ found by \citet{sub00} and intermediate between the values
advocated by \citet{nav96,nav97} of $\alpha = 1.0$ and \citet{ghi00} of
$\alpha = 1.5$. We will calculate here the lensing probability of two 
measurable variables: image separation and time delay, examining in a
subsequent paper the expected arc properties. 

Recently, \citet{dav02} used lensing statistics to constrain the inner 
slope of lensing galaxies. Using the Schechter function \citep{sch76}, 
they constrained the inner slope of lensing galaxies to the range from 
$1.58$ to $1.98$, at 95\% confidence level (CL). It is hard to predict
how their result would change if the Press-Schechter function \citep{pre74}
were used. Our choice of $\alpha = 2$ for galaxies is supported by the
following fact: stellar dynamics of elliptical galaxies, modeling of
lensed systems, and flux ratios of multiple images all give an inner
profile that is consistent with SIS 
\citep{rix97,rom99,coh01,rus01,tre02,rus02,rus03}.

\citet{san02} have reported a remarkably flat inner slope in the lensing
cluster MS2137-23: $\alpha < 0.9$ at 99\% CL. However, by measuring the 
average gravitational shear profile of six massive clusters of virial masses 
$\sim 10^{15} M_\odot$, \citet{dah02} have found that the data are well 
fitted by a mass density profile with $\alpha \sim 0.9-1.6$ for SCDM model 
and $\alpha \sim 1.3-1.6$ for LCDM model, both at 68\% CL. So, our choice
of $\alpha = 1.3$ for population B looks reasonable. The inner density
slopes for small mass halos are not well constrained. CDM simulations 
generally predict a cusped inner density, while other dark matter models,
like warm dark matter \citep{bod01}, repulsive dark matter \citep{goo00},
and collisional dark matter \citep{spe00}, tend to predict flatter
inner density (see also Ricotti 2002). Our choice of $\alpha = 1.3$ for 
population C should be
a reasonable upper limit. In her recent paper, by requiring that the 
Schechter luminosity function and the Press-Schechter mass function to 
give consistent predictions for the image separation below $1^{\prime\prime}$, 
\citet{ma03} has shown that the fraction of SIS halos peaks around mass of 
$10^{12} M_\odot$ and quickly drops for large and small mass halos. This 
is qualitatively consistent with the model that we adopt in this paper.

The paper is organized as follows: In \S\ref{sec2} we write down the
lensing cross-section produced by
SIS and GNFW halos. In \S\ref{sec3} we show how to calculate the lensing 
probability, assuming that halos are composed of the population defined 
above, whose mass function is given by the Press-Schechter function 
\citep{pre74}. In \S\ref{sec4} we present our results. In \S\ref{sec5} 
we summarize and discuss our results.

\section{Lensing by SIS and GNFW Halos}
\label{sec2}

Issues related to image separation are presented in LO02 in
detail, so here we focus on the time delay between multiple images
produced by gravitational lensing.

\subsection{Singular isothermal sphere}
\label{sec21}

The density profile for an SIS is \citep{got74,tur84}
\begin{eqnarray}
	\rho(r) = \frac{\sigma_v^2}{2\pi G} \frac{1}{r^2} \;, 
	\label{sis}
\end{eqnarray}
where $\sigma_v$ is the constant velocity dispersion. 

Assuming that the angular-diameter distances from the observer to the 
lens and the source are respectively $D_L^A$ and $D_S^A$, from the lens 
to the source is $D_{LS}^A$. Then, the time delay between the two images 
of the remote source lensed by an SIS halo is
\citep{sch92}
\begin{eqnarray}
	\Delta t = \Delta t_1\, y \;, \hspace{1cm}
		\Delta t_1 \equiv \frac{8 \pi \xi_0}{c}\left(
		\frac{\sigma_v}{c}\right)^2\, (1+ z_L) \;,
\end{eqnarray}
where 
\begin{eqnarray}
	\xi_0 \equiv 4 \pi \left(\frac{\sigma_v}{c}\right)^2 D_R^A \;,
		\hspace{1cm}
		D_R^A \equiv \frac{D_L^A D_{LS}^A}{D_S^A} \;,
\end{eqnarray}
$z_L$ is the redshift of the lens (dark halo), $y$ is the distance from the 
source to the point where the line of sight through the lens center intersects 
the source plane, in units of $\eta_0 \equiv \xi_0 D_S^A / D_L^A$\,. 

The cross-section for producing two images with a time delay $> \Delta t$ 
is
\begin{eqnarray}
	\sigma(>\Delta t) = \pi \xi_0^2 \left[1 - \left(\frac{\Delta 
		t}{\Delta t_1}\right)^2\right]\, \vartheta(\Delta 
		t_1 - \Delta t) \;,
	\label{siga}
\end{eqnarray}
where $\vartheta(\Delta t_1 - \Delta t)$ is the step function.

\subsection{Generalized NFW profile}
\label{sec22}

The density profile for a GNFW profile is (Zhao 1996; Wyithe, Turner, \&
Spergel 2001; LO02)
\begin{eqnarray}
	\rho(r) = \frac{\rho_s r_s^3}{r^\alpha (r + r_s)^{3-\alpha}} \;,
	\label{gnfw}
\end{eqnarray}
where $\alpha$, $\rho_s$, and $r_s$ are constants. The case of $\alpha =
1$ corresponds to the NFW profile \citep{nav96,nav97}. The case of 
$\alpha = 2$, $r_s \rightarrow \infty$ but keeping $\rho_s r_s^2$ constant, 
corresponds to the SIS profile. In this paper, we take $\alpha = 1.3$ for 
Populations B and C.

In the lens plane, we denote the distance from the lens center to the
point where the light ray of the source object intersects the lens plane 
by $x$, in units of
$r_s$. In the source plane, we denote the distance from the source to
the point where the line of sight through the lens center intersects
the source plane by $y$, in units of $r_s D_S^A / D_L^A$. Then, the
lensing equation is (LO02)
\begin{eqnarray}
	y = x - \mu_s \frac{g(x)}{x} \;, \hspace{0.8cm}
	g(x) \equiv \int_0^x u du \int_0^\infty dz(u^2 +
		z^2)^{-\alpha/2} \left[(u^2 + z^2)^{1/2} + 
		1\right]^{\alpha -3} \;,
\end{eqnarray}
where 
\begin{eqnarray}
	\mu_s \equiv \frac{16 \pi G}{c^2} \rho_s r_s D_R^A \;.
\end{eqnarray}
		 
To a good approximation, the time delay between the two images produced 
by a GNFW halo is given by \citep{ogu02}
\begin{eqnarray}
	\Delta t \approx \Delta t_2\, \frac{y}{y_r} \;, \hspace{1cm}
	\Delta t_2 \equiv \frac{2 r_s^2 x_t}{c D_R^A}\, 
	(1+ z_L)\, y_r \;,
\end{eqnarray}
where $x_t$ is the positive root of $y(x) = 0$, $y_r$ corresponds to the 
positive $y$ at $dy/dx = 0$.

The cross-section for producing two images with a time delay $> \Delta 
t$ is
\begin{eqnarray}
	\sigma(>\Delta t) = \pi y_r^2 r_s^2 \left[1 - \left(\frac{
		\Delta t}{\Delta t_2}\right)^2\right]\, 
		\vartheta(\Delta t_2 - \Delta t) \;.
	\label{sigb}
\end{eqnarray}

\section{Gravitational Lensing Produced by the Compound Population}
\label{sec3}

The probability for a remote point source lensed by foreground dark
halos is given by 
\begin{eqnarray}
	P = 1 - e^{-\tau} \,, \hspace{1cm}
	\tau \equiv \int_0^{z_S} dz_L\,\frac{d D_L}{dz_L} 
		\int_0^\infty d M\, n(M,z_L) \sigma(M,z_L) \;,
	\label{ip}
\end{eqnarray}
where $z_S$ is the redshift of the source, $D_L$ is the proper distance 
from the observer to a lens at redshift $z_L$, $n(M,z_L) dM$ is the
proper number density of lens objects of masses between $M$ and $M+
dM$, $\sigma(M,z_L)$ is the lensing cross-section of a dark halo of mass
$M$ at redshift $z_L$. When $\tau \ll 1$ (which is true in most cases
for lensing statistics), we have $P \approx \tau$.

For both SIS and GNFW profiles, the mass contained within radius $r$ 
diverges as $r \rightarrow \infty$. So, a cutoff in radius must be 
introduced. Here, as is typically done in the literature, we define the 
mass of a dark halo to be the mass within a sphere of radius $r = 
r_{200}$, where $r_{200}$ is the radius within which the average mass 
density is $200$ times the critical mass density of the universe at
the redshift of the halo. 

As in LO02, we consider three kinds of cosmological models: LCDM, OCDM,
and SCDM. We assume that the number density of dark halos is distributed 
in mass according to the Press-Schechter function \citep{pre74}. We 
compute the CDM power spectrum using the fitting formula given by 
\citet{eis99}, where, to be consistent with the recent observations of
{\em WMAP} \citep{spe03}, we assume the Hubble constant $h = 0.7$ and
the primordial spectrum index $n_s = 0.96$. For OCDM and SCDM, we 
determine the value of $\sigma_8$ by the cluster abundances constraint 
\citep{wan98,wan00}
\begin{eqnarray}
	\sigma_8 \Omega_m^{\gamma} \approx 0.5 \;,
	\label{s8}
\end{eqnarray}
where $\gamma \approx 0.33 + 0.35 \Omega_m$. For LCDM, we take $\Omega_m$ 
and $\sigma_8$ to be consistent with the observations of {\em WMAP} 
\citep{spe03}: $\Omega_m = 0.27$ (then $\Omega_L = 0.73$), $\sigma_8 = 
0.84$. 

A new cluster abundances constraint has recently been obtained by \citet{bah02} 
with the SDSS data. The best-fit cluster normalization is given by $\sigma_8 
\Omega_m^{0.6} = 0.33\pm 0.03$ (for $0.1 \la \Omega_m \la 0.4$) for the 
flat model of the Universe with a Hubble constant $h = 0.72$. \citet{bah02}
found that the best-fit parameters of the observed mass function are
$\Omega_m = 0.19\pm_{0.07}^{0.08}$ and $\sigma_8 = 0.9\pm_{0.2}^{0.3}$.
Recent calibration of the cluster data based on X-ray observations 
\citep{bor01,rei02,sel02,via02} are closer to the \citet{bah02} result. So,
for comparison, we will also present some results for a flat LCDM model with
the Bahcall et al. normalization to show the sensitivity of results to 
normalization.

For the case of image separation, the cross-section $\sigma$ can be 
found in LO02 (eqs.~[37] for SIS and [48] for GNFW). For the case of time 
delay, the cross-section is given by equation~(\ref{siga}) for SIS halos, 
and equation~(\ref{sigb}) for GNFW halos. 

We normalize the GNFW profile so that the concentration
parameter $c_1 \equiv r_{200}/r_s$ satisfies \citep{ogu01,ogu02}
\begin{eqnarray}
	c_1(M,z_L) = c_{\rm norm} \frac{2-\alpha}{1+z_L}\left(\frac{M}
		{10^{14} h^{-1} M_\odot}\right)^{-0.13} \;.
	\label{c1}
\end{eqnarray}
Throughout the paper we fix $c_{\rm norm} = 8$, in consistence with the
simulations \citep{bul01}.

For the model of compound halo population considered in this paper, the 
integration over mass $M$ is divided into three parts: $\int_0^{M_a}$ for 
GNFW with $\alpha = 1.3$, $\int_{M_a}^{M_b}$ for SIS, and 
$\int_{M_b}^{\infty}$ for GNFW with $\alpha = 1.3$\,; where $M_a = 10^{10} 
h^{-1} M_\odot$, $M_b = 2\times 10^{13} h^{-1} M_\odot$\,.

\section{Results}
\label{sec4}

With the formalism described above, we are ready to calculate the lensing
probability for images separation and time delay. The models to be calculated 
are listed in Table~\ref{tab1}. As explained in the previous section, we take 
three different normalizations for LCDM models: in most of calculations we 
choose parameters to be consistent with {\em WMAP} \citep{spe03}, but, for 
comparison, we will also present some results corresponding to the normalization
of \citet{bah02}.  For OCDM and SCDM models, we adopt equation~(\ref{s8})
for normalization. Throughout the paper we take $h = 0.7$ and $n_s = 0.96$.

\subsection{Image separation}
\label{sec4.1}

For image separation, we have calculated the differential lensing 
probability
\begin{eqnarray}
	\frac{d P}{d \lg \Delta\theta} \equiv
		- \frac{d P(>\Delta\theta)}{d \lg \Delta\theta}\;,
\end{eqnarray}
where $P(> \Delta\theta)$ is given by equation~(\ref{ip}) with $\sigma 
= \sigma (>\Delta\theta)$. We show the results for different cosmological
models in Figure~\ref{fig1}, 
separately for the three different components in the whole population: 
Population A (galaxies, the highest island), Population B (groups and
clusters of galaxies, the second high island), and Population C (dwarf
galaxies and subgalactic objects, the lowest island). The source object 
is assumed to be at $z_S = 3$.

From the figure we see that, Population A (galaxies) contributes most to 
the total number of lenses, due to its steep inner density slope ($\alpha 
= 2$); Population B contributes less; Population C contributes least, due 
to its small mass and shallow inner density slope ($\alpha = 1.3$).  
Consistent with the results in LO02, the lensing probability 
produced by the $\alpha =1.3$ GNFW halos is smaller than the lensing
probability produced by SIS halos by two orders of magnitudes in the 
overlap regions. (The results here are slightly different from those in 
LO02 due to the fact that in this paper we use a different normalization 
in the concentration parameter, i.e. eq.~[\ref{c1}].)

In Figure~\ref{fig2}, we show the LCDM ($\Omega_m = 0.27$, $\sigma_8 =
0.84$) results corresponding to different redshift of the source object:
from $z_S = 1$ to $z_S = 10$. We see that the lensing probability increases
quickly with the source redshift, increasing by an order of 
magnitude between $z_S =1$ and $z_S =3$ (cf. Wambsganss, Bode, \& Ostriker 
2003). However, the rate of increase in the
lensing probability decreases with the source redshift, this is because 
that the proper distance from the source object to the observer approaches 
a finite limit as $z_S \rightarrow \infty$ (due to the existence of a 
horizon in an expanding universe). We also see that, as the source redshift 
increases, the splitting angle corresponding to the peak probability of 
each island shifts toward larger values. In Figure~\ref{fig3}, we show the 
corresponding integral lensing probability 
\begin{eqnarray}
	P(<\Delta\theta) = P(>0) -P(>\Delta\theta) \;.
\end{eqnarray}

To compare the predictions with observations, the effect of magnification 
bias must be considered (Turner, Ostriker, \& Gott 1984; Schneider, Ehlers, 
\& Falco 1992; LO02; Oguri et al. 2002). When the source
objects have a flux distribution $\propto f^{-\beta}$ ($\beta <3$) and the 
probability density for magnification is $\propto A^{-3}$, the magnification 
bias is given by (LO02)
\begin{eqnarray}
	B = \frac{2}{3-\beta} A_m^{\beta -1} \;,
	\label{bias}
\end{eqnarray}
where $A_m$ is the minimum of the total amplification. For SIS lenses we 
have $A_m = 2$. For GNFW lenses, $A_m$ can be approximated by
\begin{eqnarray}
	A_m \approx \frac{2x_t}{y_r y^\prime_t} \;,
	\label{amnfw}
\end{eqnarray}
where $y^\prime_t \equiv (dy/dx)(x = x_t)$. 

Equation~(\ref{amnfw}) is an improvement to the equation (68) of LO02. The
magnification bias calculated with equations~(\ref{bias}) and (\ref{amnfw})
agrees with that calculated with the more complicated formula of 
\citet{ogu02} with errors $<5\%$ for $0.01 <\mu_s < 10$.

For GNFW lenses with $\alpha = 1.3$, we show the average magnification bias 
$\langle B\rangle$ (defined by the ratio of the biased lensing probability 
to that without bias) as a function of image separation in Figure~\ref{fig4a} 
(as an improvement to the Fig.~10 of LO02) for the JVAS/CLASS survey 
\citep{mey03,bro02}, where we have assumed $\beta = 2.1$ \citep{rus01b} and $z_S 
= 1.27$ \citep{mar00}. The magnification bias for GNFW lenses depends on 
cosmological models, decreases with increasing image separation, and is
bigger than the magnification bias for SIS lenses (which is a constant
$B \approx 4.76$) by about $1.2$ order of magnitude on average (for $\alpha
= 1.3$).

In Figure~\ref{fig4}, we compare our predictions (including magnification bias)
for the compound model 
with observations from the JVAS/CLASS survey. The data are updated compared 
to \citet{hel00}. The new data contain $13$ lenses found in a sample of
$8958$ of radio sources which form a statistical sample \citep{bro02}. 
Considering error bars, both LCDM and OCDM models with both normalizations
are marginally consistent with the JVAS/CLASS observational 
data.\footnote{The apparent excess in the prediction for small splitting
angles is caused by the angular selection effect: the JVAS/CLASS survey
is limited to image separation $\ge 0.3^{\prime\prime}$ \citep{bro02}.} 
Comparing
LCDM2 with LCDM3, we find that even for the same cluster normalization 
there is significant discriminatory power available from lensing statistics
(if data is available) in breaking the degeneracy on the $(\Omega_{\rm m},
\sigma_8)$ plane. This is consistent with our previous results (LO02).
The three different LCDM models do not differ significantly in their 
predictions at small splittings but for splittings above 10 arcseconds
the Bahcall et al normalization, LCDM3, predicts few lenses by more than a
factor of five.

\subsection{Time delay}
\label{sec4.2}

For time delay, we have calculated the differential lensing probability
\begin{eqnarray}
	\frac{d P}{d \lg \Delta t} \equiv
		- \frac{d P(>\Delta t)}{d \lg \Delta t}\;,
\end{eqnarray}
where $P(> \Delta t)$ is given by equation~(\ref{ip}) with $\sigma 
= \sigma (>\Delta t)$. We show the results in Figure~\ref{fig5}, for 
the same models in \S\ref{sec4.1}.

We see that, the distribution of lensing probability over time delay
is very similar to the distribution over image separation (compare
Fig.~\ref{fig2} to Fig.~\ref{fig1}). Again, the contribution to 
lensing events is overwhelmingly dominated by Population A due to its
steep inner density slope. Population C contributes the least.

We have also calculated the lensing probability for time delay corresponding
to different source redshift: from $z_S = 1$ to $z_S = 10$. 
The results for the LCDM model ($\Omega_m = 0.27$, $\sigma_8 = 0.84$) are 
shown in Figure~\ref{fig6} for the differential lensing probability $dP/d\lg 
\Delta t$, and Figure~\ref{fig7} for the integral lensing probability 
\begin{eqnarray}
	P(<\Delta t) = P(>0) - P(>\Delta t) \;. 
\end{eqnarray}
From these figures we see that, like in the case for image separation, the
lensing probability sensitively depends on $z_S$ for small $z_S$. For large
$z_S$, the lensing probability becomes less sensitive to the source 
redshift, due to the fact that $dD_S/dz_S$ decreases with increasing
$z_S$. 

We can calculate the joint lensing probability $P(>\Delta\theta,\, > 
\Delta t)$ by using the joint cross-section
\begin{eqnarray}
	\sigma(>\Delta\theta,\, > \Delta t) = \sigma(>\Delta\theta)\,
		\vartheta(\Delta t_i - \Delta t) \;,
\end{eqnarray}
where $\Delta t_i = \Delta t_1$ for SIS and $\Delta t_2$ for GNFW.
The cross-section $\sigma(>\Delta\theta)$ is given by equation~(37) of LO02
for SIS, and equation~(48) of LO02 for GNFW. Then, we can calculate the 
conditional lensing probability $P(\Delta\theta,\, > \Delta t)$ defined by
\begin{eqnarray}
	P(\Delta\theta,\, > \Delta t) \equiv - \frac{\partial}
		{\partial\Delta\theta} P(>\Delta\theta,\, > 
		\Delta t) \;,
	\label{pst}
\end{eqnarray}
which gives the distribution of lensing events over time delay for
a given image separation.

Knowing $P(\Delta\theta,\, > \Delta t)$, we can calculate the median
time delay $\Delta t_{\rm med}$ as a function of $\Delta\theta$, 
where $\Delta t_{\rm med}$ is defined by
\begin{eqnarray}
	P(\Delta\theta,\, > \Delta t_{\rm med}) = \frac{1}{2}
		P(\Delta\theta,\, > 0) \;.
	\label{med}
\end{eqnarray}
The prediction for $\Delta t_{\rm med}$ as a function of $\Delta\theta$ is
not sensitive to the magnification bias since it is determined by the 
ratio of two probabilities. So, the correlation between $\Delta t_{\rm 
med}$ and $\Delta\theta$ provides a test of lensing models independent of
the determination of magnification bias.

The results of $\Delta t_{\rm med}(\Delta\theta)$ for the LCDM ($\Omega_m 
= 0.27$, $\sigma_8 = 0.84$; indeed the results are insensitive to 
the cosmological parameters) model are shown in Figure \ref{fig8}, where 
the source object is again assumed to be at $z_S = 1.27$. In 
Figure~\ref{fig8} we also show the quadrant deviations (dashed lines), which 
are defined by equation~(\ref{med}) with the $1/2$ on the right-hand side being 
replaced by $1/4$ and $3/4$, respectively. The observational data, taken
from \citet{ogu02}, fit the LCDM model well. Comparison of Figure~\ref{fig8} 
with \citet{ogu02}'s Figure~6 indicates that our compound model fits the 
observations better. The single population model predicts a single (almost)
straight line in the $\lg\Delta\theta - \lg\Delta t_{\rm med}$ space. For 
the compound model, a ``step'' is produced at the point where the mass density
profile changes. The ``step'' that we see in Figure~\ref{fig8} corresponds 
to the transition from population A (galaxies) to population B (galaxy 
groups/clusters).

\section{Summary and Discussion}
\label{sec5}

\begin{figure}
\plotone{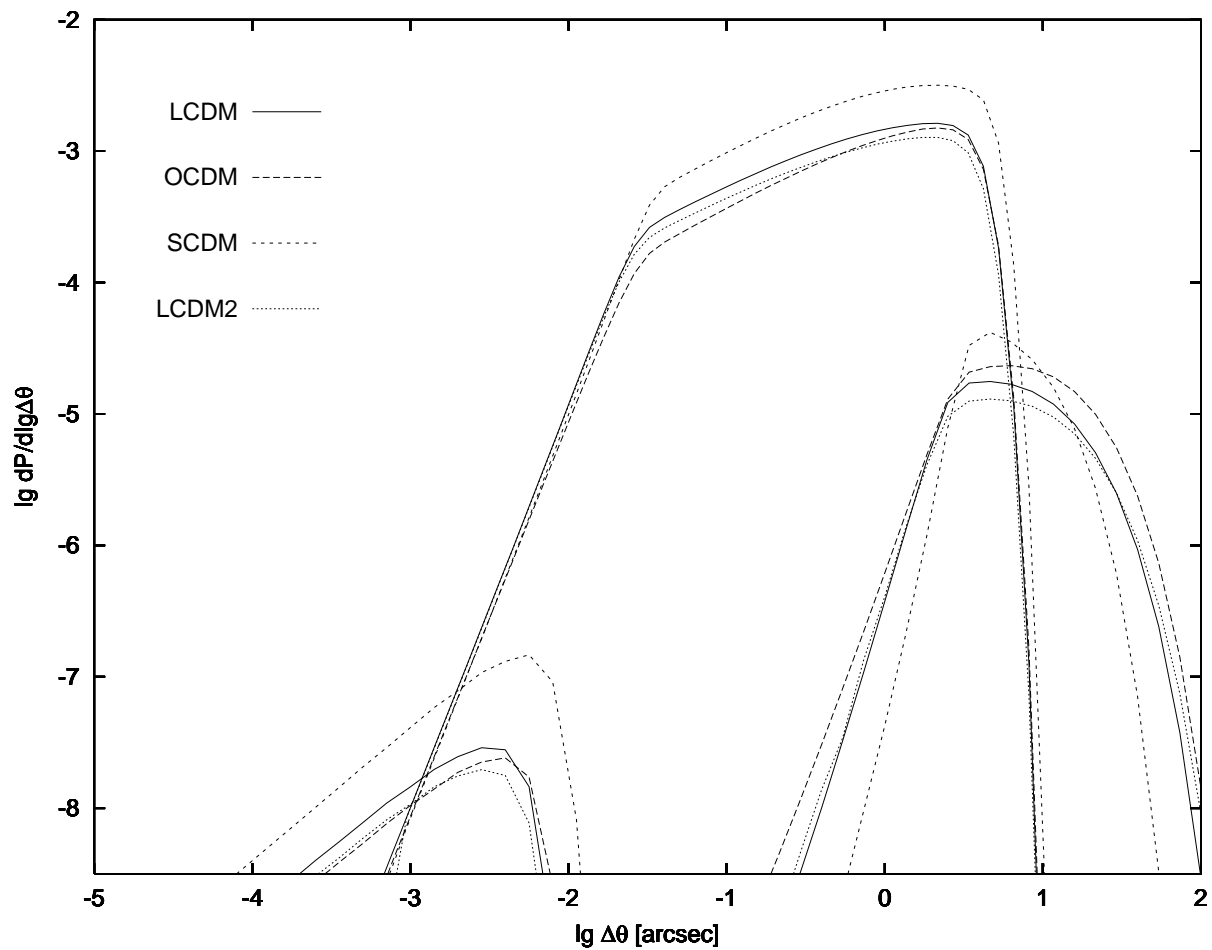}
\caption{Differential lensing probability of image separation for the 
three components in the compound population of dark halos. Four cosmological
models are shown: LCDM, OCDM, SCDM, and LCDM2 (for definitions see 
Table~\ref{tab1}). The tallest 
islands in the middle are for Population A (galaxies). The lower islands on 
the right are for Population B (cluster halos). The smallest islands on the 
left are for Population C (dwarf halos). The source object is at $z_S = 3$. 
\label{fig1}}
\end{figure}

As an extension of our previous work (LO02), we computed the lensing
probability produced by a compound population of dark halos. 
We have calculated the lensing probability for both image separation and
time delay. The calculations confirm our previous results (LO02) that the
lensing probability produced by GNFW halos with $\alpha \la 1.3$ is lower
than that produced by SIS halos with same masses by orders of magnitudes,
where $-\alpha$ is the inner slope of the halo mass density. So, for the
compound population of halos, both the number of lenses with large image 
separation ($\Delta\theta \ga 5^{\prime\prime}$) and the number of lenses 
with small image separation ($\Delta\theta \la 10^{-2\,\prime\prime}$)
are greatly suppressed. The same conclusion holds also for the number of
lenses with large time delay ($\Delta t \ga 10\,{\rm years}$) and the number 
of lenses with small time delay ($\Delta t \la 10^{-4}\,{\rm year}$).
(See Figs.~\ref{fig1} and \ref{fig5}. This conclusion holds even when 
the effect of magnification bias is considered, see Figs.~\ref{fig4a} and 
\ref{fig4}.)

\begin{figure}
\plotone{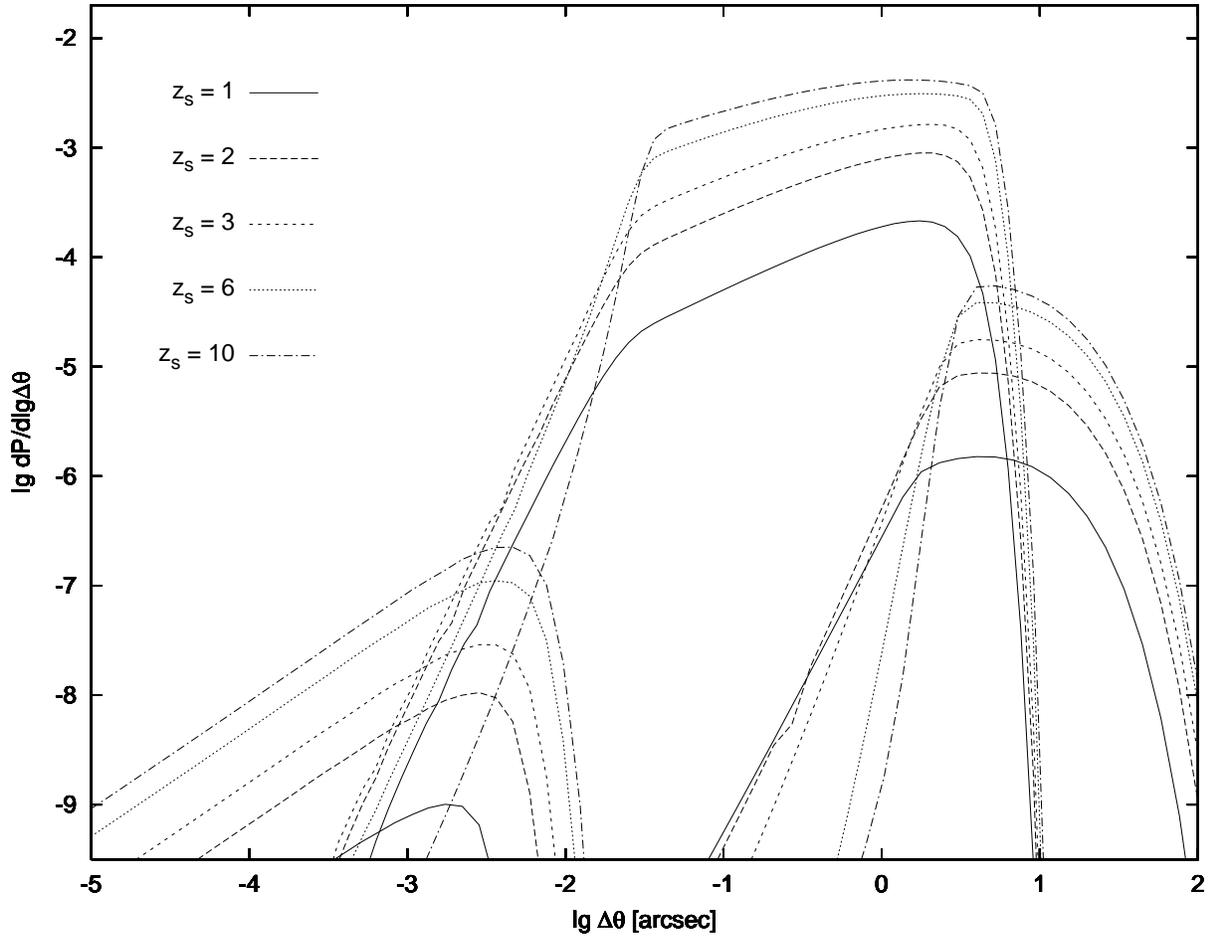}
\caption{Differential lensing probability of image separation for the 
LCDM model. Different type of lines corresponds to 
different redshift of the source object as indicated in the figure. 
As in Fig.~\ref{fig1}, each island in the figure corresponds to one 
of the three components in the compound population: Population A (center), 
Population B (right), and Population C (left). 
\label{fig2}}
\end{figure}

\begin{figure}
\plotone{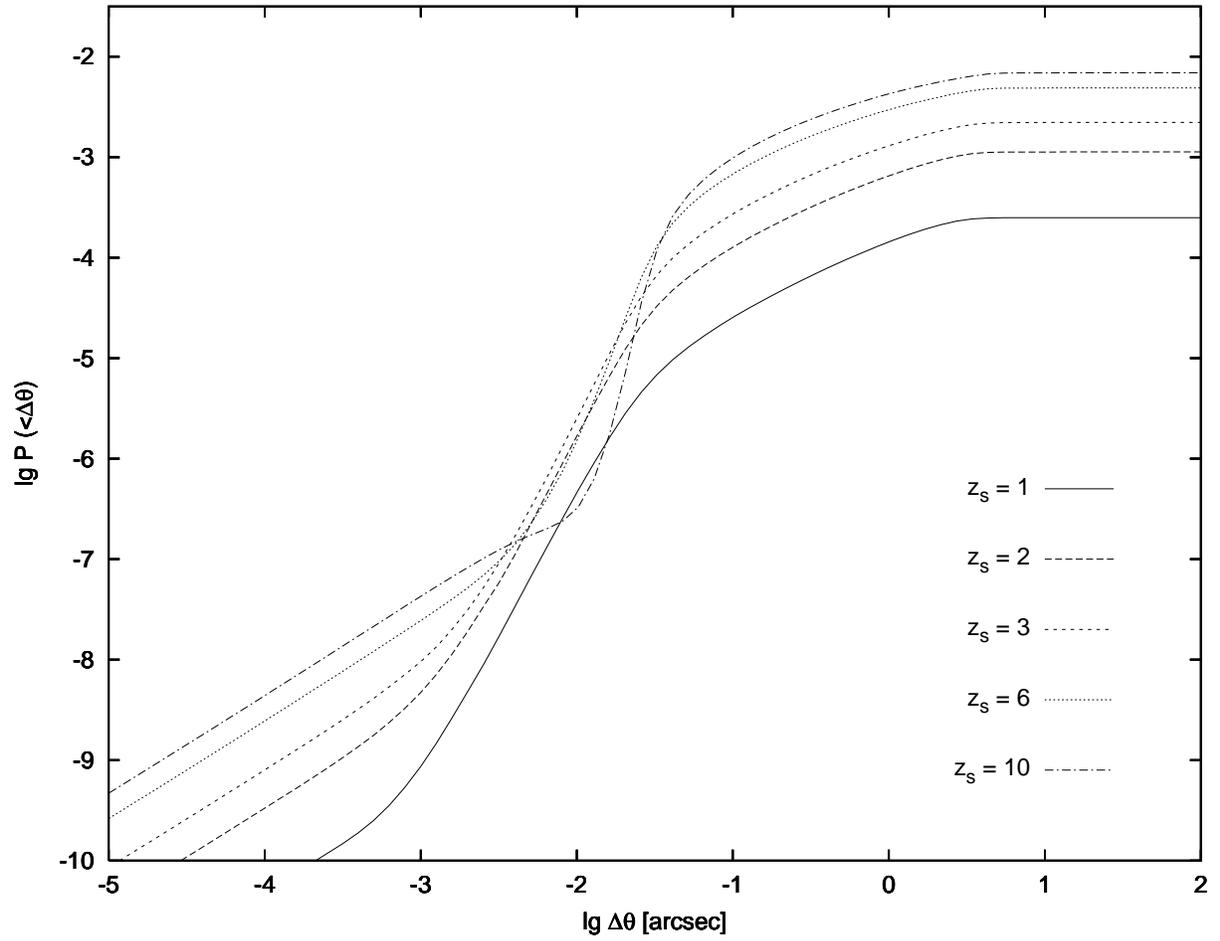}
\caption{The integral lensing probability $P(<\Delta \theta)$ produced by 
the compound population of dark halos. The models and the meaning of 
symbols are the same as those in Fig.~\ref{fig2}.
\label{fig3}}
\end{figure}

\begin{figure}
\plotone{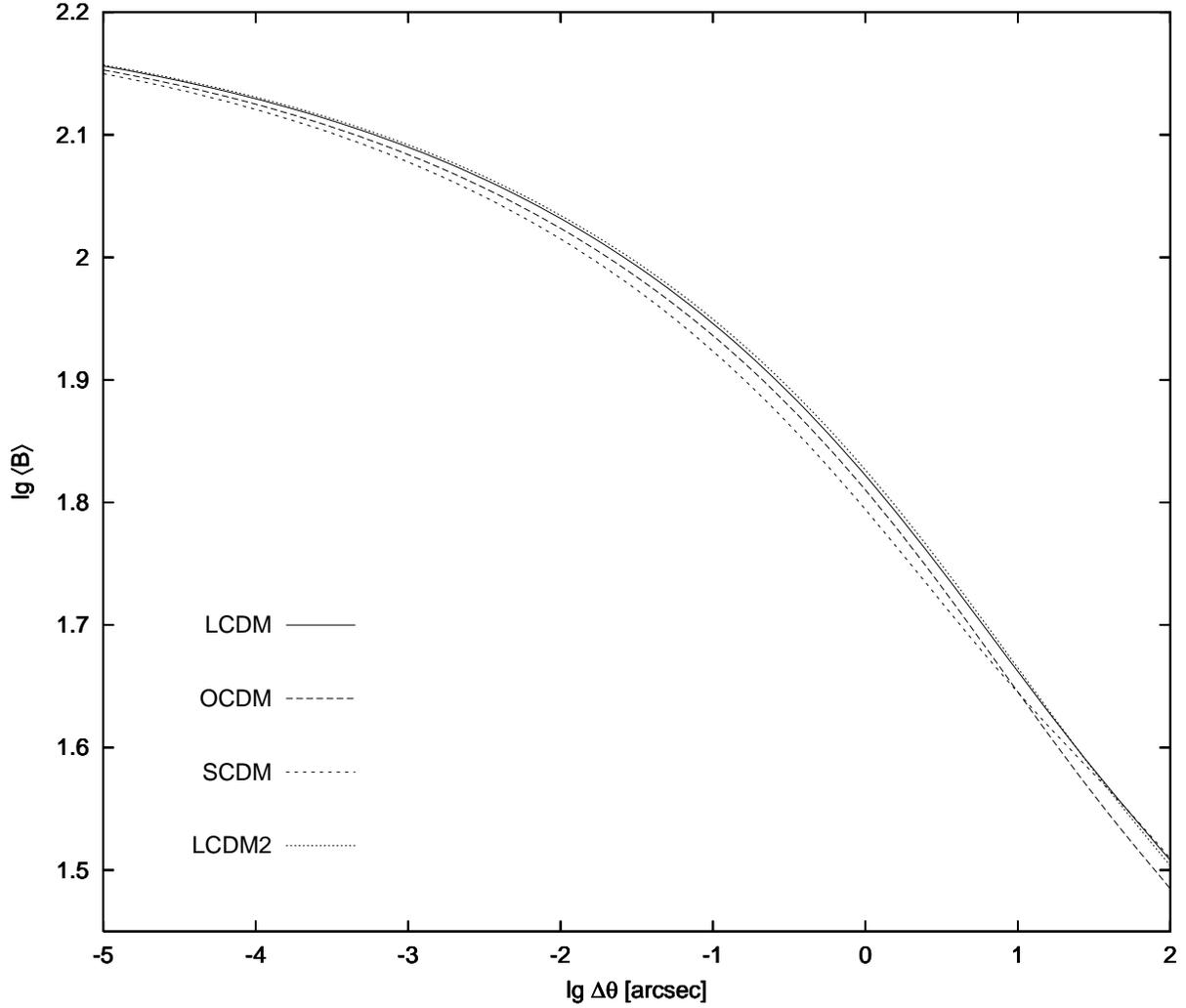}
\caption{Average magnification bias as a function of image separation for
GNFW lenses with $\alpha = 1.3$ and $\beta = 2.1$. The cosmological models 
and the meaning of symbols are the same as those in Fig.~\ref{fig1}, except 
that $z_S = 1.27$ (to be consistent with the JVAS/CLASS survey). (For 
comparison, the corresponding magnification bias for SIS lenses is a 
constant $B = 4.76$.)
\label{fig4a}}
\end{figure}

\begin{figure}
\plotone{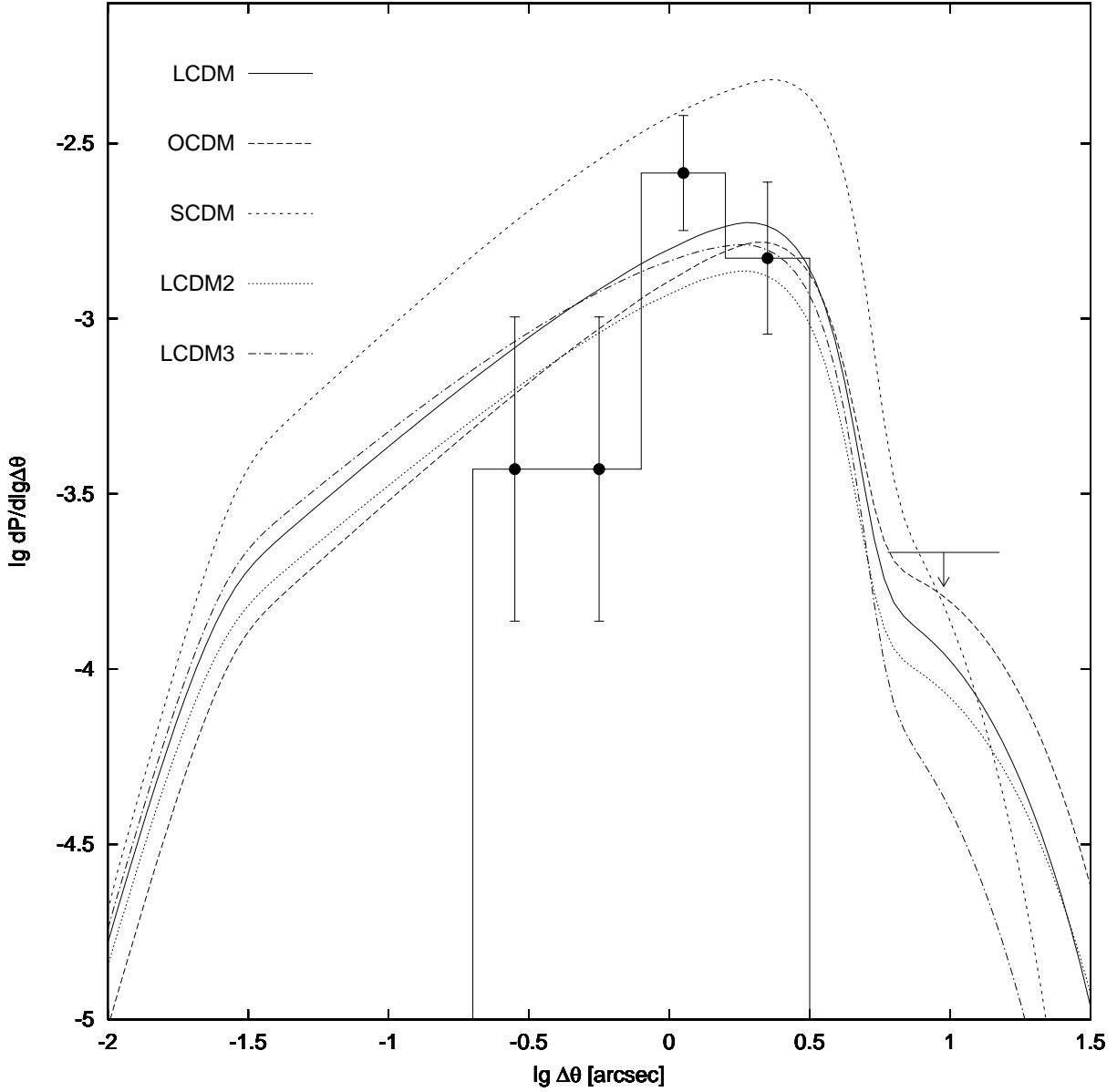}
\caption{Comparison with the JVAS/CLASS data. Predictions by five different
cosmological models are shown with different type of curves as indicated
(see Table~\ref{tab1} for definitions for the cosmological models).
The redshift of the source object is assumed to be $z_S = 1.27$. The data
with error bars are taken from JVAS/CLASS survey \citep{bro02}. The null
result for lenses with $6^{\prime\prime} \le \Delta\theta \le 15^{\prime
\prime}$ is shown with the horizontal line with a downward arrow indicating 
that is an upper limit.
\label{fig4}}
\end{figure}

\begin{figure}
\plotone{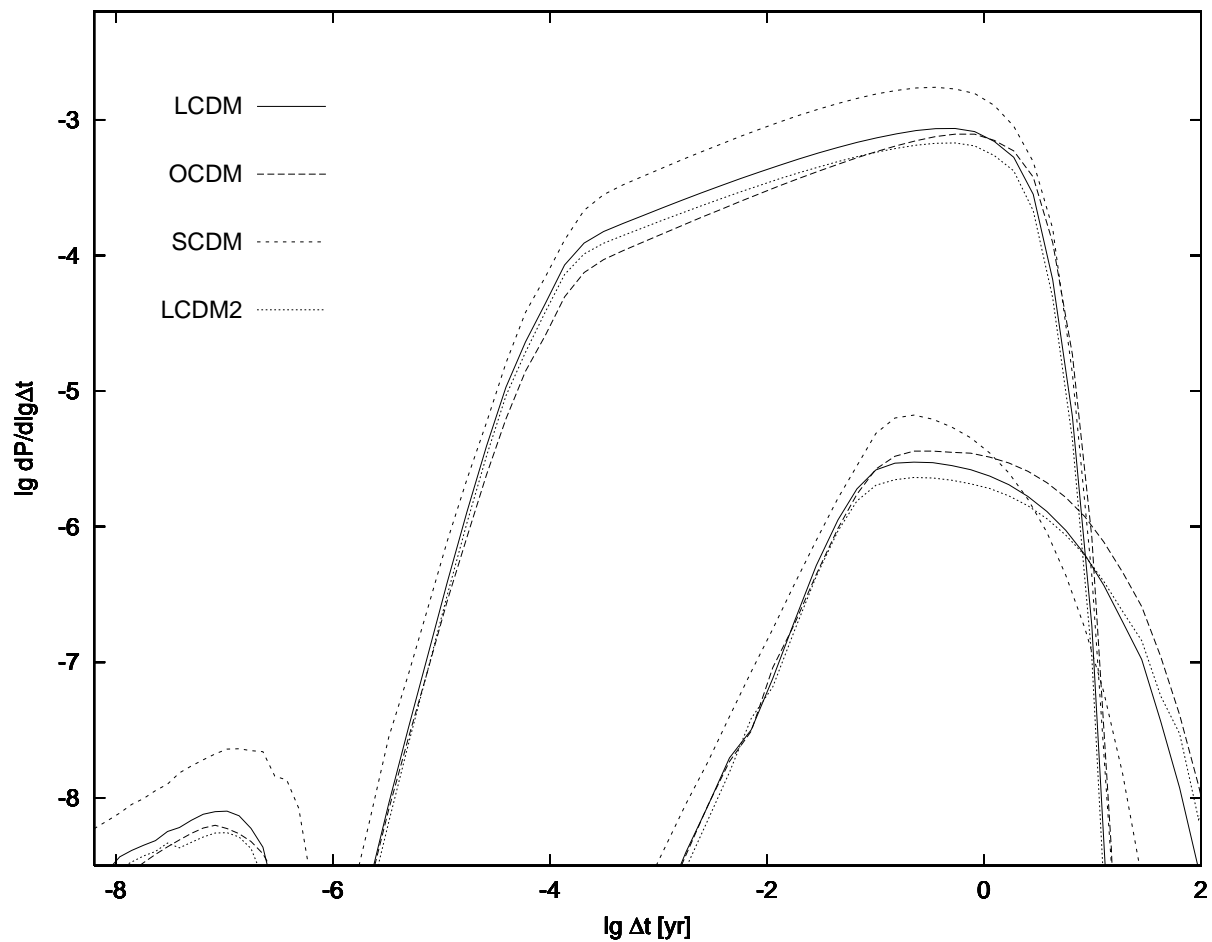}
\caption{Differential lensing probability of time delay for the three 
components in the compound population of dark halos. The models and
the meaning of the symbols are the same as those in Fig.~\ref{fig1}.
Of the studied models only SCDM can be excluded by the JVAS/CLASS
observational data (see Fig.~\ref{fig4}).
\label{fig5}}
\end{figure}

\begin{figure}
\plotone{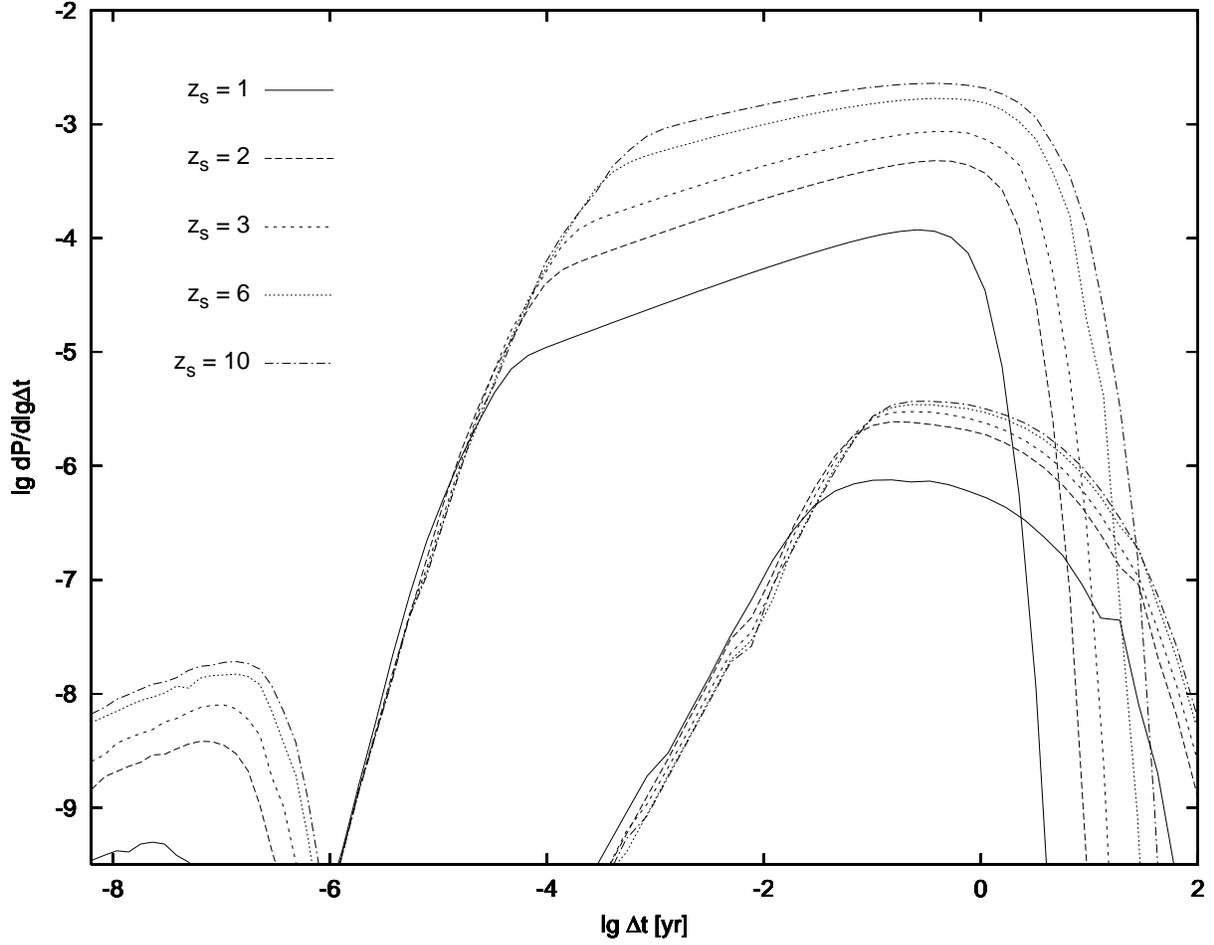}
\caption{Differential lensing probability of time delay for the LCDM 
model corresponding to different 
redshift of the source object. The models and the meaning of the symbols 
are the same as those in Fig.~\ref{fig2}.
\label{fig6}}
\end{figure}

\begin{figure}
\plotone{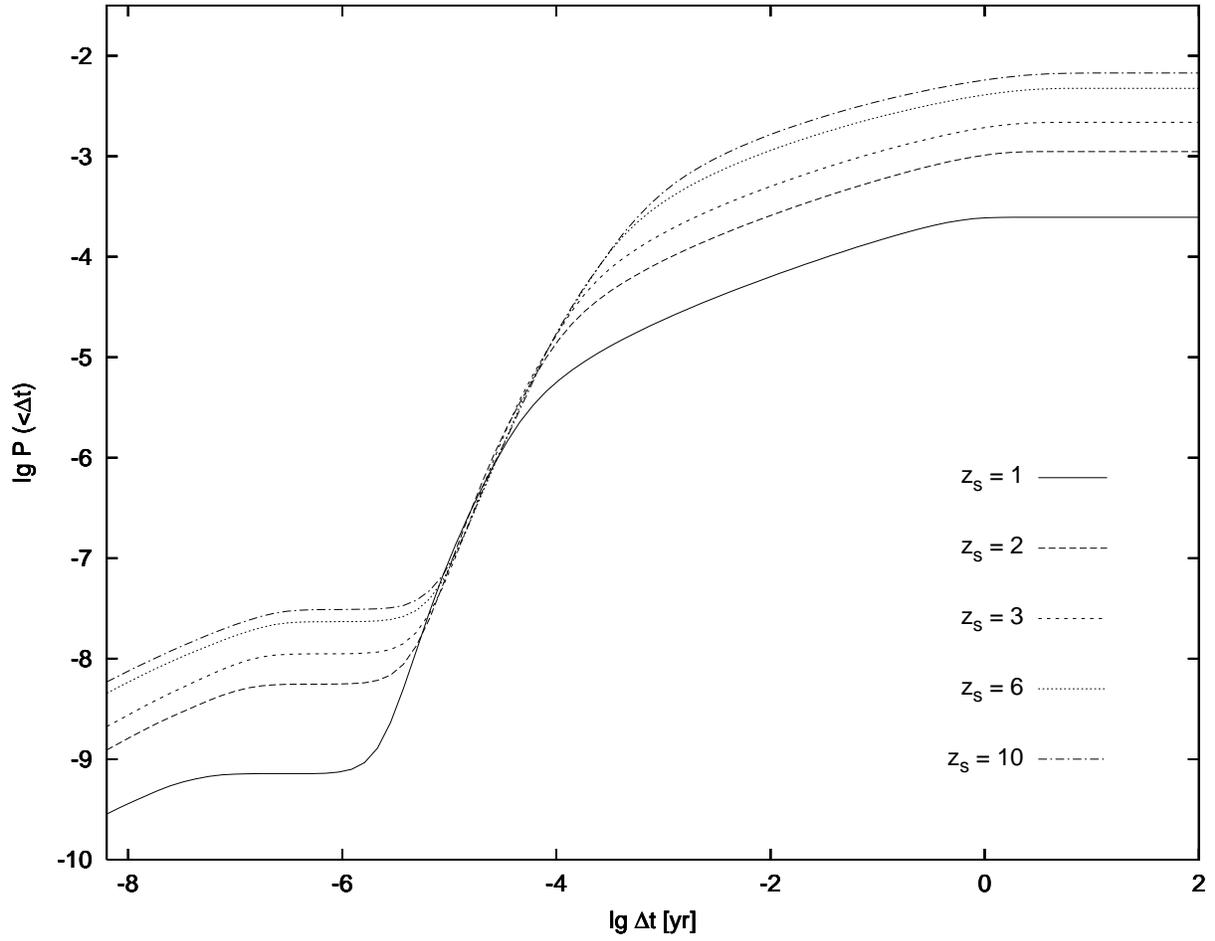}
\caption{The same as Fig.~\ref{fig6} but for the integral lensing 
probability $P(<\Delta t)$.
\label{fig7}}
\end{figure}

We have also tested the dependence of the lensing probability on the 
redshift of the source object (Figs.~\ref{fig2}, \ref{fig3}, \ref{fig6}, and 
\ref{fig7}). The results show that, the lensing probability is quite sensitive 
to the change in the redshift of the source object. The number of lenses 
significantly increases as the source redshift increases. However, the rate 
of the increase decreases as the source redshift becomes large, which is 
caused by the fact  
that the proper cosmological distance approaches a finite limit when $z_S
\rightarrow \infty$. Another interesting result is that, the peak of the 
lensing probability for each population moves toward large image separation
or time delay, as the source redshift increases. 

We see that population C (dwarf halos) in an LCDM model has a unique 
signature in the time domain, c.f. Figures~\ref{fig5} and \ref{fig6}. Time
delays of less than $10$ seconds and greater than 0.1 second are predicted
and should be found in gamma-ray burst sources which are at cosmological
distances and have the requisite temporal substructure. Variants of CDM,
such as warm dark matter \citep{bod01}, repulsive dark matter \citep{goo00},
or collisional dark matter \citep{spe00} would not produce this feature.
However, current surveys do not go deep enough to provide a sufficiently 
large sample to test the prediction. When more observational 
data on gamma-ray burst time delay and small splitting angles become
available, our calculations can be used to distinguish different dark
matter models \citep{nem01,wil01}.

We have compared the distribution of the number of lenses over image 
separation predicted by our model with the updated JVAS/CLASS observational
data, with the new {\em WMAP} cosmological parameters (Fig.~\ref{fig4}). 
Since the JVAS/CLASS survey is limited to image 
separation $\ge 0.3^{\prime\prime}$ \citep{bro02}, we cannot test our
predictions for small image separations. However, in the range that is probed
by JVAS/CLASS, we see that both the LCDM and OCDM models fit the observation
reasonably well and current data do not allow us to distinguish between the
two proposed normalizations for the LCDM spectrum, even though these produce
predictions that differ by a factor of roughly $1.4$. 
An explicit search for lenses with image separation between 
$6^{\prime\prime}$ and $15^{\prime\prime}$ has found no lenses \citep{phi01}, 
which rules out the SIS model for image separation in this range (LO02). This
together with our Figure~\ref{fig4} supports our model of compound population
of halos. For separations greater than $10^{\prime\prime}$ the differently 
normalized 
LCDM models produce significantly different results, thus producing an
additional lever to break the degeneracies in the WMAP results (cf. Bridle
et al. 2003)

\begin{figure}
\plotone{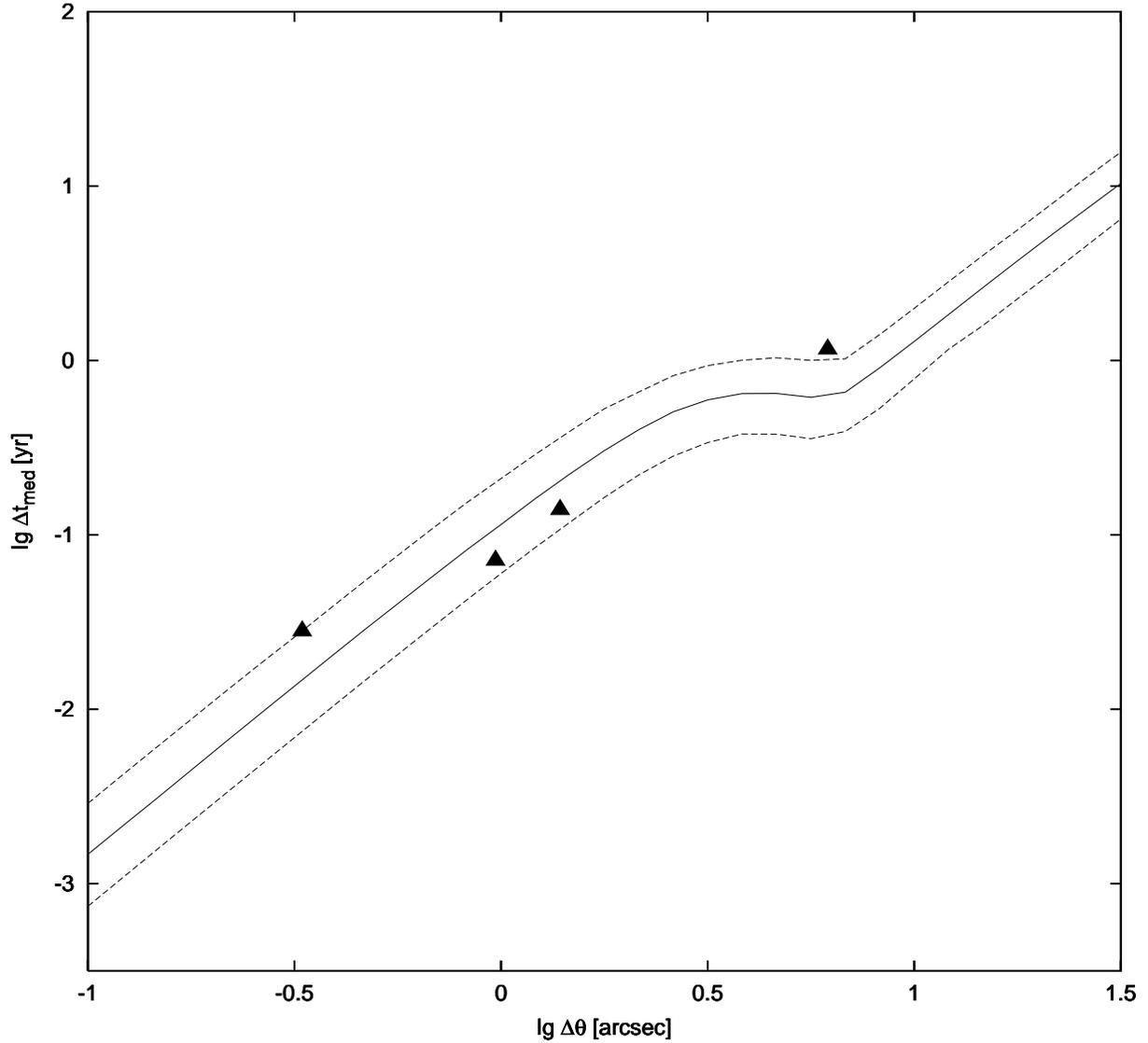}
\caption{The median distribution of time delay (solid line) as a function 
of image separation, produced by the compound model of dark halos in the 
LCDM model. (The results for this figure are insensitive to the cosmological 
parameters.) The source object is assumed to be at $z_S = 1.27$. The dashed
lines show the quadrant deviations. The observational data are taken from
\citet{ogu02}.
\label{fig8}}
\end{figure}

We have also calculated the distribution of the mean time delay vs image
separation for the LCDM model (Fig.~\ref{fig8}). We see that, the compound 
model fits observations quite well, better than the model of single population 
of halos \citep{ogu02}. The compound model predicts a unique feature in the 
$\lg\Delta\theta - \lg\Delta t_{\rm med}$ plane: there is a ``step'' 
corresponding to the transition in mass density profile. This can be better 
tested when more observation data are available.

A controlled survey of lenses with double the sample size of CLASS, perhaps
obtainable via SDSS \citep{yor00,sto02}, should allow one to better 
distinguish between LCDM variants and perhaps between LCDM models and those 
based on quintessence \citep{cal98} rather than a cosmological constant.

\acknowledgments

We thank B. Paczy\'{n}ski for many helpful discussions, and the anonymous
referee whose comments helped to improve our results.
LXL's research was supported by NASA through Chandra Postdoctoral Fellowship 
grant number PF1-20018 awarded by the Chandra X-ray Center, which is operated 
by the Smithsonian Astrophysical Observatory for NASA under contract 
NAS8-39073. JPO's research was supported by the NSF grants ASC-9740300 
(subaward 766) and AST-9803137.

\clearpage
\begin{deluxetable}{lllllll}
\tablewidth{0pt}
\tablecaption{Cosmological Models Calculated in the Paper\label{tab1}}
\tablehead{
\colhead{Model} & \colhead{$\Omega_m$} & \colhead{$\Omega_\Lambda$} &
\colhead{$\sigma_8$} & \colhead{$h$} & \colhead{$n_s$}& \colhead{Normalization} 
}
\startdata\vspace{0.2cm}
LCDM  & ~$0.27$~ & ~$0.73$~ & ~$0.84$~ & ~$0.7~$ & ~$0.96$~ & {\it WMAP\tablenotemark{a}}   \\
LCDM2 & ~$0.19$~ & ~$0.81$~ & ~$0.9$~ & ~$0.7$~ & ~$0.96$~ & $\sigma_8 \Omega_m^{0.6} = 0.33$\tablenotemark{b} \\
LCDM3 & ~$0.3$~ & ~$0.7$~ & ~$0.7$~ & ~$0.7$~ & ~$0.96$~ & $\sigma_8 \Omega_m^{0.6} = 0.33$\tablenotemark{b} \\ 
OCDM  & ~$0.3$~ & ~$0$~ & ~$0.85$~ & ~$0.7$~ &~$0.96$~ & $\sigma_8 \Omega_m^{\gamma} = 0.5$\tablenotemark{c} \\ 
SCDM  & ~$1$~ & ~$0$~ & ~$0.5$~ & ~$0.7$~ & ~$0.96$~ & $\sigma_8 \Omega_m^{\gamma} = 0.5$\tablenotemark{c}
\enddata
\tablenotetext{a}{From \citet{spe03}.}
\tablenotetext{b}{From \citet{bah02}.}
\tablenotetext{c}{From \citet{wan98} and \citet{wan00}, $\gamma =
0.33 + 0.35 \Omega_m$.}

\tablecomments{See the text for meanings of symbols.}

\end{deluxetable}

\clearpage

\clearpage

\clearpage

\clearpage

\clearpage

\clearpage

\clearpage

\clearpage

\clearpage

\end{document}